\documentclass[reprint,aps,pre,amsmath,amssymb,superscriptaddress]{revtex4-1}
\usepackage{graphicx,hyperref}
\hypersetup{colorlinks=true, linkcolor=blue, citecolor=blue, urlcolor=blue}

\renewcommand{\v}[1]{\mathbf{#1}}
\newcommand{\gv}[1]{\mbox{\boldmath$ #1 $}}
\newcommand{\uv}[1]{\mathbf{\hat{#1}}}
\newcommand{\abs}[1]{\left| #1 \right|}

\renewcommand{\d}[0]{\ensuremath{\operatorname{d}\!}}

\let\f=\frac

\renewcommand{\l}[0]{\left}
\renewcommand{\r}[0]{\right}
\newcommand{\fp}[2]{\l( \f{#1}{#2} \r)}

\newcommand{\ie}[0]{\textit{i.e.}}
\newcommand{\eg}[0]{\textit{e.g.}}

\renewcommand{\t}[1]{\text{#1}}

\newcommand{\Ang}[0]{\, \mathring{\mathrm{A}}}
\newcommand{\kB}[0]{k_\t{B}}
\newcommand{\K}[0]{\, \t{K}}
\newcommand{\ns}[0]{\, \t{ns}}

\newcommand{\avg}[1]{\langle #1 \rangle}
\newcommand{\Avg}[1]{\l< #1 \r>}

\begin{document}
  \title{Capillary fluctuations of surface steps: An atomistic simulation study for the model Cu(111) system}
  \author{Rodrigo Freitas}
  \email{rodrigof@berkeley.edu}
  \affiliation{Department of Materials Science and Engineering, University of California, Berkeley, CA 94720, USA}
  \affiliation{Lawrence Livermore National Laboratory, Livermore, California 94550, USA}
  \author{Timofey Frolov}
  \affiliation{Lawrence Livermore National Laboratory, Livermore, California 94550, USA}
  \author{Mark Asta}
  \affiliation{Department of Materials Science and Engineering, University of California, Berkeley, CA 94720, USA}
  \date{\today}
  \begin{abstract}
    Molecular dynamics (MD) simulations are employed to investigate the capillary fluctuations of steps on the surface of a model metal system. The fluctuation spectrum, characterized by the wave number ($k$) dependence of the mean squared capillary-wave amplitudes and associated relaxation times, is calculated for $\avg{110}$ and $\avg{112}$ steps on the $\{111\}$ surface of elemental copper near the melting temperature of the classical potential model considered. Step stiffnesses are derived from the MD results, yielding values from the largest system sizes of $(37\pm1) \, \t{meV}/\Ang$ for the different line orientations, implying that the stiffness is isotropic within the statistical precision of the calculations. The fluctuation lifetimes are found to vary by approximately four orders of magnitude over the range of wave numbers investigated, displaying a $k$ dependence consistent with kinetics governed by step-edge mediated diffusion. The values for step stiffness derived from these simulations are compared to step free energies for the same system and temperature obtained in a recent MD-based thermodynamic-integration (TI) study [Freitas, Frolov, and Asta, Phys. Rev. B {\bf 95}, 155444 (2017)]. Results from the capillary-fluctuation analysis and TI calculations yield statistically significant differences that are discussed within the framework of statistical-mechanical theories for configurational contributions to step free energies.
  \end{abstract}
  \maketitle

  \section{\label{sec:introduction} Introduction}
    Capillary fluctuations are a ubiquitous phenomenon at fluid interfaces, line defects, and crystalline interfaces that are atomically rough \cite{fluidfluid,cwm_prl,solidsolid,bartelt_1,lv_1,lv_2,macdowell}. These equilibrium fluctuations, which lead to variations in the line length of a linear defect, or area of a rough interface at finite temperature, have been widely studied by advanced experimental characterization techniques and computer simulations, as they provide insights into the thermodynamic and kinetic properties of the interfaces on which they form. While a detailed overview of such studies is beyond the scope of the present manuscript, we refer the reader to comprehensive reviews and representative experimental and computational studies \cite{williams_2,einstein,muller_review,bartelt_2,k_scaling_williams,williams_3} in the context of steps at faceted crystalline interfaces, which provide the focus of the present work. The properties of such steps play a critical role in governing the kinetics of crystal growth from melt, solution, or vapor phases, due to their influence on the thermodynamics of island nucleation and the kinetics of interface migration (\eg, Ref.~\cite{ice}).

    Over the last decade analyses of capillary fluctuations in molecular-scale computer simulations, based on molecular dynamics (MD) or Monte Carlo (MC) methods, have been employed extensively within the so-called capillary-fluctuation method (CFM) approach to computing interfacial free energies and their associated crystalline anisotropies for crystal-melt interfaces, grain boundaries, and solid-solid heterophase interfaces (\eg, Refs.~\cite{cwm_prl,cm,cm_2,k_scaling_karma,cwm_original,solidsolid,gb,gb_2}). Recently, the CFM approach has been employed also for steps at faceted crystal-melt interfaces \cite{peyman} to derive temperature-dependent step stiffnesses, which are relevant in the context of modeling solidification rates and associated crystal growth morphologies. For crystal-melt interfaces, liquid surfaces, and fluid-fluid interfaces, detailed comparisons of CFM results with those obtained using alternative thermodynamic-integration and nucleation based MD methods have been undertaken to understand the range of applicability and associated accuracies of these alternative approaches (\eg, Refs.~\cite{grest_1,grest_2,grest_3,laird_1,laird_2,laird_3,lj_morris}). At the present time we are unaware of such comparisons for the applications of the CFM for step properties.

    In the present work we consider the application of the CFM approach for studying thermodynamic and kinetic properties of steps on crystalline surfaces, focusing on Cu$(111)$ as a representative model metal system. The results of equilibrium MD simulations near the melting temperature of the potential model considered are analyzed to compute step fluctuation spectra, characterized by the wave number ($k$) dependence of the mean-square amplitudes $\avg{\abs{A(k)}^2}$, as well as the fluctuation relaxation times $\tau(k)$. From the dependence of $\avg{\abs{A(k)}^2}$ on $k$ we derive step stiffnesses for $\avg{110}$ and $\avg{112}$ step orientations, obtaining values that are isotropic (independent of orientation) within the statistical precision of the simulations. Further, we obtain values of fluctuation lifetimes that are consistent with the $k^{-4}$ scaling associated with dynamics that are governed by step-edge diffusion.

    The focus on the Cu$(111)$ system in the present study enables a comparison of CFM results with step free energies obtained in a recent study published by the authors \cite{freitas_prb} using an alternative thermodynamic-integration (TI) approach. The values of the step stiffnesses derived by the CFM are lower by approximately $25$\% compared with the step free energies calculated by the TI approach for $\avg{110}$ oriented steps at the same temperature. The discrepancy is discussed within the framework of statistical-mechanical theories of the configurational contributions to step free energies (\eg, Refs.~\cite{fisher,kayser}) associated with capillary fluctuations.

    The remainder of this paper is organized as follows. In Sec.~\ref{sec:theory} we present a brief derivation of the main results from capillary-wave theory that are used in the remainder of the paper; although similar derivations appear already in many places in the literature, the overview is included to emphasize key concepts and equations required for the analysis and interpretation of the present MD results. In Sec.~\ref{sec:simulations} we describe the details of the MD simulations of step capillary fluctuations and in Sec.~\ref{sec:results} we present the simulation results. In Sec.~\ref{sec:discussion} a discussion is presented focusing on the comparison of the present CFM results to step free energies obtained previously by thermodynamic integration \cite{freitas_prb}. Finally, the results and conclusions are summarized in Sec.~\ref{sec:summary}.

  \section{\label{sec:theory} Capillary-wave model}
    In this section we summarize the main equations required for CFM analysis of surface step fluctuations. We describe how the capillary-wave Hamiltonian results from the coarse-graining of the atomic partition function, and also highlight several nuances of the CFM that will be discussed in the context of the analysis of the MD simulation results in Sec.~\ref{sec:simulations}.

    \subsection{\label{sec:theory_1} Step effective Hamiltonian}
\begin{figure}
        \centering
        \includegraphics[width=0.45\textwidth]{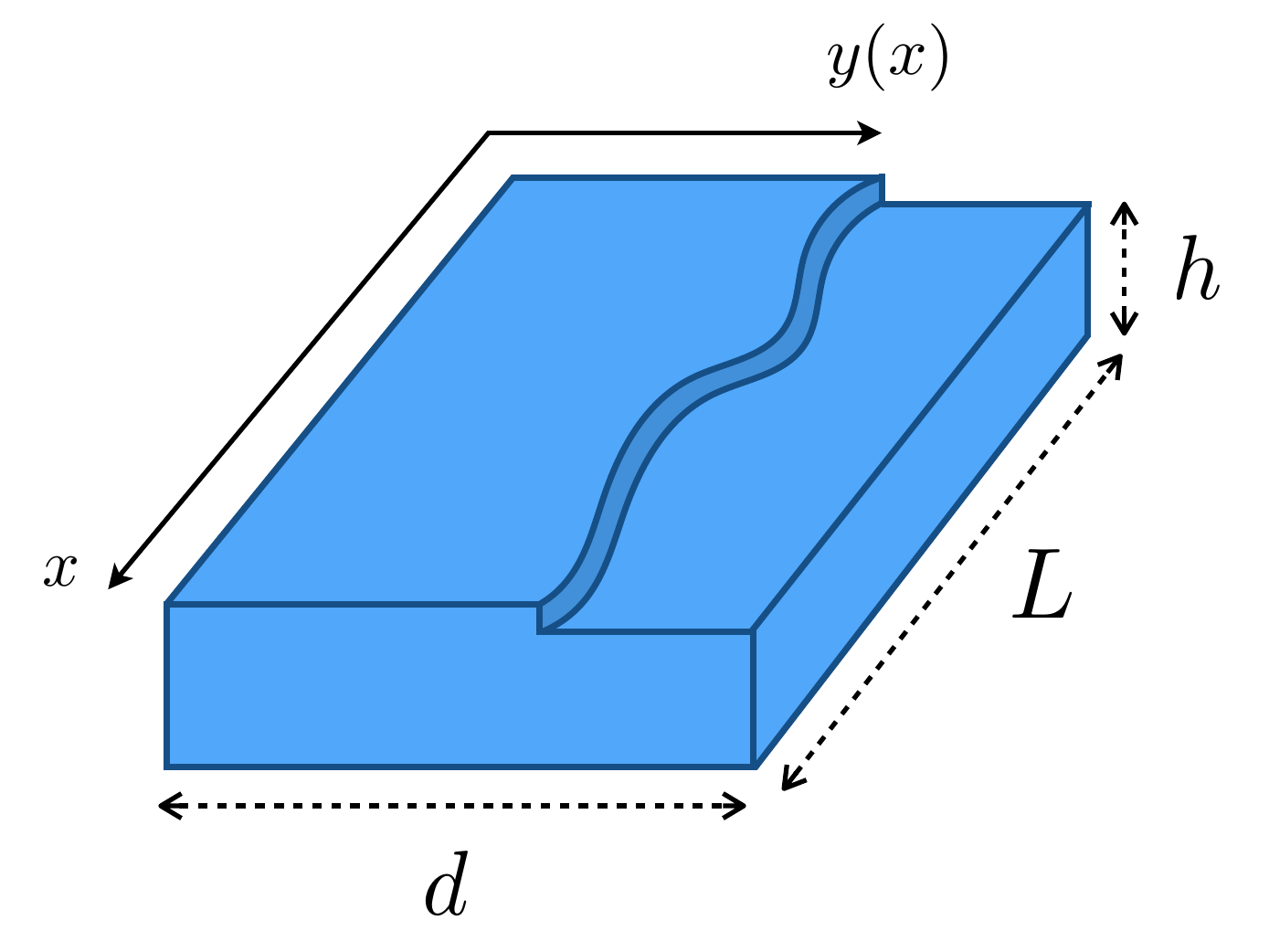}
        \caption{System dimensions and orientation: $d$ is the step-step separation distance, $L$ is the step length, and $h$ is the bulk depth. The step average direction is along $\uv{x}$, and the step line profile is given by the curve y(x). Periodic boundary conditions are applied on the $x$ and $y$ directions.}
        \label{fig:step_illustration}
      \end{figure}

      Following the notation from Ref.~\cite{freitas_prb}, the excess free energy of a step can be defined thermodynamically through the relation:
      \begin{equation}
        \label{eq:thermo_F}
        [ F ]_{AN} \equiv F^\t{st} - F^\t{t} = \gamma^\t{st} L,
      \end{equation}
      where $\gamma^\t{st}$ is the step free energy per unit length and $L$ is the system dimension along the average step direction, as illustrated in Fig.~\ref{fig:step_illustration}. $F^\t{st}$ and $F^\t{t}$ are the absolute free energies of systems with the same surface area ($A$), number of atoms ($N$), and temperature ($T$). These systems can be considered to be identical except that the system corresponding to $F^\t{t}$ has a flat surface, while the one corresponding to $F^\t{st}$ contains a surface step of length $L$. The free energy of the system with a flat surface can also be written as $F^\t{t} = - \kB T \ln Q^\t{t} + 3N \kB T \ln \Lambda$, where $\kB$ is the Boltzmann constant, $Q^\t{t}$ is the configurational part of the system's partition function, and $\Lambda = (h^2/ 2\pi m \kB T)^{1/2}$ is the thermal de Broglie wavelength. Similarly, the free energy of the system with a step is $F^\t{st} = - k_\t{B} T \ln Q^\t{st} + 3N \kB T \ln \Lambda$. Hence, we can rewrite the step free energy in Eq.~\eqref{eq:thermo_F} as follows:
      \begin{equation}
        \label{eq:fs_Q}
        [ F ]_{AN} = - k_\t{B} T \ln Q,
      \end{equation}
      where $Q \equiv Q^\t{st}/Q^\t{t}$ is the ratio of the configurational partition functions.

      In order to clarify the physical meaning of $Q$ consider the potential energy of the system with a step: $U^\t{st}(\v{x})$, where $\v{x}$ is the $3N$-dimensional vector with the atomic coordinates. We can perform a canonical transformation on the atomic coordinates and separate the variables describing the step configuration from all other variables. With this transformation the potential energy can be written as $U^\t{st} = U^\t{st}(\v{R},\v{r})$, where $\v{R}$ are the step degrees of freedom and $\v{r}$ represents all other degrees of freedom (\ie, bulk and surface degrees of freedom). With this set of generalized coordinates the configurational partition function of the system with the step can be written as
      \begin{equation}
        \label{eq:cg_Z}
        Q^\t{st} = \int \d{ \v{R}} \, \exp\bigg[-\beta U_\t{cg}(\v{R})\bigg],
      \end{equation}
      where
      \begin{equation}
        \label{eq:cg_H}
        U_\t{cg}(\v{R}) = - k_\t{B} T \ln \l\{ \int \d{\v{r}} \, \exp\bigg[-\beta U^\t{st}(\v{R}, \v{r})\bigg] \r\} ,
      \end{equation}
      is a coarse-grained potential energy which involves only the step degrees of freedom. Notice that, for convenience, we have performed the canonical transformation in such a way as to render $\v{R}$ and $\v{r}$ dimensionless quantities. Equation \eqref{eq:cg_H} implies that $U_\t{cg}(\v{R})$ is the portion of the free energy associated with the bulk and surface configurational degrees of freedom $\v{r}$. Alternatively, Eq.~\eqref{eq:cg_Z} suggests that $U_\t{cg}(\v{R})$ can also be seen as the potential that generates the step dynamics on that system. Because of this last interpretation $U_\t{cg}(\v{R})$ is also known as the potential of mean force \cite{tuckerman}, \ie, it is the potential acting on the step that arises from the mean contribution of the bulk and surface degrees of freedom. In the limit of adiabatic decoupling between the step and the rest of the system $U_\t{cg}(\v{R})$ becomes an effective potential on which the step degrees of freedom ($\v{R}$) can be assumed to evolve in time independently from the other degrees of freedom ($\v{r}$).

      The step free energy can be written as a function of the coarse-grained potential energy $U_\t{cg}(\v{R})$. First, we substitute Eq.~\eqref{eq:cg_Z} in the equation for $Q$ and use $Q^\t{t} = \exp\l[-\beta (F^\t{t}-3N\kB T \ln \Lambda)\r]$:
      \[
        Q = \f{Q^\t{st}}{Q^\t{t}}
          = \int \d{\v{R}} \, \exp \bigg[ -\beta \mathcal{H}(\v{R}) \bigg],
      \]
      where we have defined the step effective Hamiltonian $\mathcal{H}(\v{R}) \equiv U_\t{cg}(\v{R}) - (F^\t{t}-3N\kB T \ln \Lambda)$. Now the step free energy can be obtained from Eqs.~\eqref{eq:thermo_F} and \eqref{eq:fs_Q}:
      \begin{equation}
        \label{eq:F_H}
        \gamma^\t{st} L = - \kB T \ln \bigg\{ \int \d{\v{R}} \, \exp \big[ -\beta \mathcal{H}(\v{R}) \big]\bigg\}.
      \end{equation}
      Notice that Eq.~\eqref{eq:F_H} does not involve any approximation, we have only separated and interpreted specific parts of the partition function $Q$. Hence, the calculation of the step free energy $\gamma^\t{st}$ using Eq.~\eqref{eq:F_H} still involves an integral over the phase space of all particles.

    \subsection{\label{sec:theory_2} Capillary-wave model for steps}
      It is now possible to introduce a model for the step effective Hamiltonian, $\mathcal{H}(\v{R})$, that simplifies the calculation of Eq.~\eqref{eq:F_H} but still includes all relevant physical properties that govern the step dynamics. A reasonable model that forms the basis for capillary-wave theory (\eg, Ref.~\cite{fisher}), is to assume that a fluctuation of the step line that causes a change $\delta \ell$ in step length has an energetic cost of $\sigma \delta \ell$, where $\sigma$ is the step energy per unit length. With this physical picture the step effective Hamiltonian takes the form:
      \begin{equation}
        \label{eq:cwm_hamilt}
        \mathcal{H} \l[ y(x) \r] = \int_y \sigma(\theta) \, \d{\ell}
                   = \int_0^L \sigma(\theta) \sqrt{1 + y'(x)^2} \d{x},
      \end{equation}
      where $\theta(x) = \tan^{-1} (y/x)$ is the step orientation with respect to the average step-line direction and the integral is over the curve $y(x)$ describing the step-line profile, as illustrated in Fig.~\ref{fig:step_illustration}. Thus, $\mathcal{H}$ is a functional of the step configuration $y(x)$ \cite{kardar}.

      Notice that $\sigma(\theta)$ defined in Eq.~\eqref{eq:cwm_hamilt} is different from the step stress tensor $\gv{\tau}^\t{st}$ as defined in, for example, Refs.~\cite{freitas_prb,step_stress}. The step stress tensor couples mechanically to the system strain and gives origin to a elastic deformation energy which can be directly measured in atomistic simulations \cite{freitas_prb}. The physical interpretation of $\sigma(\theta)$ is more complicated, as discussed in detail in Ref.~\cite{fisher}. For example, $\sigma(\theta)$ reflects the energy per unit \textit{physical length} of the step, while $\gv{\tau}^\t{st}$ and $\gamma^\t{st}$ are defined per unit length of the average step direction, indicated as $L$ in Fig.~\ref{fig:step_illustration} and Eq.~\eqref{eq:thermo_F}. For our purpose in this paper we will refer to $\sigma(\theta)$ as the step tension in the line-fluctuation model given by Eq.~\eqref{eq:cwm_hamilt}.

      To compute from Eq.~\eqref{eq:cwm_hamilt} the equilibrium spectrum for the capillary fluctuations, and the resulting free energy, the traditional approach \cite{cwm_original,fisher,kardar} is to make use of the small slope approximation where $\theta(x) \approx y'(x)$ and the terms inside the integral in Eq.~\eqref{eq:cwm_hamilt} can be expanded in powers of $y'(x)$. Collecting the terms with the same power and keeping only terms $O(y'^2)$ allow us to write the step effective Hamiltonian as 
      \begin{equation}
        \label{eq:Hs_1}
        \mathcal{H}[y(x)] = \sigma L + \f{1}{2} \tilde{\sigma} \int_0^{L} y'(x)^2 \d{x},
      \end{equation}
      where $\sigma \equiv \sigma(0)$ is the step tension of the state with a straight step in this model, and $\tilde{\sigma} \equiv \sigma(0) + \sigma''(0)$ is the step stiffness, where $\sigma''(0)$ denotes the second derivative of the step tension with respect to the orientation of the step normal evaluated in the state where the step is straight. The term $\sigma L$ in Eq.~\eqref{eq:Hs_1} is the energy of a straight step, while the second term is the energy penalty in having any curvature along the step line, \ie, the energy cost of step fluctuations.

      The next step is to discretize the integral in Eq.~\eqref{eq:Hs_1} into a Riemann sum, resulting in a Hamiltonian that is quadratic in $y(x_n)$, with $x_n = n \Delta x$ where $n=0, 1, \ldots, M-1$ and $\Delta x = L / M$. In what follows we adopt a similar approach based on a Fourier representation of the step profile. This formulation, while equivalent, leads to expressions more aligned with the CFM analysis of computer simulation results.

      The step line profile $y(x)$ shown in Fig.~\ref{fig:step_illustration} can be decomposed in normal modes as:
      \begin{equation}
        \label{eq:fourier}
        y(x) = \sum_{n=-(M-1)/2}^{(M-1)/2} A_n \exp\l(ik_nx\r),
      \end{equation}
      where the wavevectors $k_n$ are given by $k_n = n (2\pi/L)$ with $n = 0, \pm 1, \pm 2, ..., \pm(M-1)/2$ (assuming $M$ is odd). Using Eq.~\eqref{eq:fourier} we can compute the integral in Eq.~\eqref{eq:Hs_1} and obtain:
      \begin{equation}
        \label{eq:hf}
        \mathcal{H}( \{A_n\} ) = \sigma L + \tilde{\sigma} L \sum_{n=1}^{(M-1)/2} k_n^2 \abs{A_n}^2 ,
      \end{equation}
      where we have made use of the fact that $A^*_n = A_{-n}$ since $y(x)$ is real. In this system of coordinates the amplitudes $A_n$ of the normal modes are the step degrees of freedom since they define the step configuration $y(x)$ through Eq.~\eqref{eq:fourier}. The step effective Hamiltonian given by Eq.~\eqref{eq:hf} is quadratic in all its degrees of freedom and thus many properties of the system can be obtained exactly. 

      It is clear from Eq.~\eqref{eq:hf} that the properties of this system depend on how the step is coarse-grained, \ie, how closely spaced ($\Delta x = L / M$) are the $M$ points describing the step line. This is a reflection of the number of degrees of freedom attributed to the step effective Hamiltonian \cite{fisher,kayser}, Eq.~\eqref{eq:cwm_hamilt}, as discussed in Sec.~\ref{sec:theory_1}; $\Delta x$ determines the largest wavevector considered in the effective Hamiltonian of Eq.~\eqref{eq:hf}: $k_\t{max} \equiv \pi / \Delta x$. An extensive review of the consequences and interpretations of this dependency on $\Delta x$ and, consequently, on the capillary-wave wavelengths considered, is given in Ref.~\cite{fisher}. We return to this point in Sec.~\ref{sec:discussion} when comparing the results of the CFM analysis to the values of the step free energy computed in Ref.~\cite{freitas_prb}.

      Notice that the term $\sigma L$ in Eq.~\eqref{eq:hf} is a simple shift in energy, thus the dynamics of the step is completely parametrized by the step stiffness $\tilde{\sigma}$. In the next section we review how $\tilde{\sigma}$ can be derived from atomistic simulations.

    \subsection{\label{sec:theory_3} Step fluctuation spectrum}
      According to the equipartition theorem each quadratic degree of freedom in the Hamiltonian of a system at constant temperature $T$ contributes $\kB T/2$ to the system's average energy. We can apply this theorem to Eq.~\eqref{eq:hf} since each mode amplitude $A_n$ appears quadratically in the Hamiltonian. Notice that the real and imaginary parts of $A_n$ are independent and, thus, each part contributes with $\kB T / 2$ to the total energy. Hence
      \[
        \tilde{\sigma} L k_n^2 \Avg{\abs{A_n}^2} = k_\t{B} T,
      \]
      where $\avg{\ldots}$ indicates a canonical ensemble equilibrium average. This equation can be used to compute the step stiffness, $\tilde{\sigma}$, if we rewrite it as
      \begin{equation}
        \label{eq:cwm}
        \Avg{\abs{A_n}^2} = \fp{\kB T}{\tilde{\sigma} L} \f{1}{k_n^2}.
      \end{equation}
      The normal-mode amplitudes can be obtained from atomistic simulations and used to adjust a curve of $\avg{\abs{A_n}^2} \; \t{versus} \; k_n^{-2}$, from which $\tilde{\sigma}$ can be extracted.

      When using Eq.~\eqref{eq:cwm} to compute $\tilde{\sigma}$ it is necessary to define the step profile, shown as $y(x)$ in Fig.~\ref{fig:step_illustration}. Hence, the value of $\tilde{\sigma}$ obtained can be sensitive to how $y(x)$ is determined, particularly if one relies on normal modes with wavelengths comparable to the atomic spacing. The physical origin of this arises because the distinction between interface and bulk degrees of freedom is not clear for atomistic systems \cite{fisher}, \ie, the definition of the interface position from the atomic configuration is ambiguous. In Sec.~\ref{sec:simulations_profile} we study the inherent ambiguity in defining the step configuration in atomic-scale simulations and discuss how $y(x)$ can be determined in such a way as to minimally affect the value of $\tilde{\sigma}$ obtained from Eq.~\eqref{eq:cwm}.

  \section{\label{sec:simulations} Methodology of atomistic simulations}
    \subsection{Molecular dynamics simulations and system geometry}
      Molecular dynamics simulations of surface steps were performed using the LAMMPS \cite{lammps} (Large-scale Atomic/Molecular Massively Parallel Simulator) software. The interatomic interactions were described by the embedded-atom method \cite{eam} for a system of pure copper \cite{Cu_Mishin} and the Langevin thermostat \cite{langevin} was used to sample the particles' phase space according to the canonical ensemble distribution. The thermostat relaxation time was $\tau_\t{L} \equiv m / \gamma = 2 \, \t{ps}$, where $\gamma$ is the friction parameter and $m$ the atomic mass. The timestep ($\Delta t$) for the integration of the equations of motion was chosen based on the phonon spectrum of the system; we used $\Delta t = 2\, \t{fs}$ which is approximately $1/60$th of the oscillation period of the highest-frequency normal mode of this system.

      The geometry of the simulation box is illustrated in Fig.~\ref{fig:step_illustration}. Periodic boundary conditions were used for the directions parallel to the surface ($\uv{x}$ and $\uv{y}$) while free boundaries were used along $\uv{z}$ to create the system surface. We kept the box length along the $\uv{x}$ and $\uv{y}$ directions fixed while the system fluctuates freely along $\uv{z}$ (normal to the surface) in order to guarantee mechanical equilibrium with the vacuum. A non-orthogonal simulation box \cite{srolovitz_steps} was employed in such a way as to have only one step on the system surface.

\begin{figure}
        \includegraphics[width=0.48\textwidth]{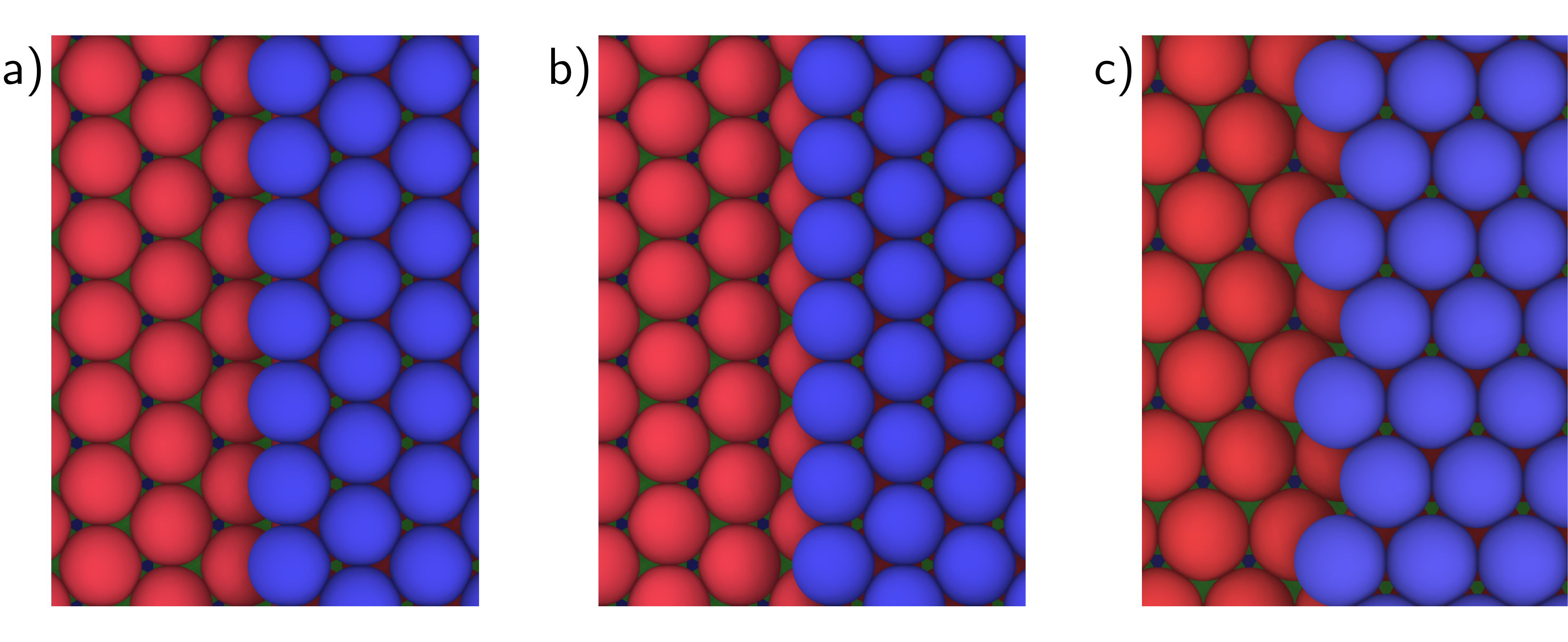}
        \caption{\label{fig:step_directions}Step orientations on the $(111)$ surface of face-centered cubic copper. The blue (dark gray) atoms on the right are in the first layer, and the red ones (light gray) on the left are in the second layer. The orientations are (a) $[110]$A, (b) $[110]$B, and (c) $[211]$. The difference between the $[110]$ A and B steps are the nearest neighbors on the layer immediately below the step. }
      \end{figure}

      We have chosen the $(111)$ surface of face-centered cubic copper as a representative metal surface to study steps. The surface properties of the interatomic potential employed have been studied extensively previously \cite{prb_tim,freitas_prb}. Of relevance for this study is the fact that the $\{111\}$ surfaces have been found to be faceted at all temperatures up to the melting point of this model ($T_\t{m} = 1327\K$). In order to study effects of anisotropy we will consider steps of two different orientations (Fig.~\ref{fig:step_directions}), with step line directions along $[110]$ and $[211]$. The $[110]$ direction presents two distinct steps, $[110]$A and $[110]$B, which have different nearest-neighbor configurations on the layer immediately below the step, as illustrated in Fig.~\ref{fig:step_directions}. The temperature for all simulations was $T = 1300\K$; it was chosen to be close to the melting point so that the step fluctuation timescales were compatible with the short physical times accessible to the MD simulations. In Sec.~\ref{sec:simulations_error} we present an analysis of the step fluctuation relaxation times to assess which modes are adequately sampled.

    \subsection{System dimensions}
      Surface steps deform the crystalline lattice around them creating an elastic field that can interact with other elastic fields present in the crystal. The total energy of an isolated step will be referred to as the self-energy ($U_0$). Because of the periodic boundary conditions applied in the simulations the step can interact with its periodic images, as well as with the surface at the bottom of the simulation box (Fig.~\ref{fig:step_illustration}). The effects of these interactions can be made negligibly small by choosing system dimensions such that the interaction energy is much smaller than $U_0$.

      The step-step interaction energy decreases with the distance $d$ between the steps as \cite{muller_review} $E_\t{int} \propto d^{-2}$ and can be attractive or repulsive depending on the orientation of the steps. To determine the magnitude of this interaction we followed the approach from Ref.~\cite{srolovitz_steps} and computed the step-step interaction energy for different step separations. We have chosen the step-step separation distance for our simulations as the minimum distance such that $E_\t{int} / U_0 \le 10^{-4}$. In practice this resulted in distances $d \approx 60 \Ang$ as shown in Table~\ref{tab:simulation_box_sizes}.
      \begin{table}
        \caption{\label{tab:simulation_box_sizes}Simulation box size used for each step orientation. $d$ is the step-step distance, $h$ the bulk depth, $L$ the step length, and $U_0$ is the step self-energy. The box geometry is illustrated in Fig.~\ref{fig:step_illustration}.}
        \begin{ruledtabular}
        \begin{tabular}{c c c c c}
          & \multicolumn{1}{c}{$d \; (\mathring{\mathrm{A}})$} & \multicolumn{1}{c}{$h \; (\mathring{\mathrm{A}})$} & \multicolumn{1}{c}{$L \; (\mathring{\mathrm{A}})$} & \multicolumn{1}{c}{$U_0 \, (\t{meV}/\mathring{\mathrm{A}})$} \\
          \hline
          $[110]$A      & $60.5$ & $42.8$ & $104.8$ & $103.1$ \\
          $[110]$B      & $66.6$ & $42.8$ & $104.8$ & $104.1$ \\
          $[211]\;\;\,$ & $62.9$ & $42.8$ & $108.9$ & $120.5$ \\
        \end{tabular}
        \end{ruledtabular}
      \end{table}

      The bulk depth ($h$ in Fig.~\ref{fig:step_illustration} and Table~\ref{tab:simulation_box_sizes}) was determined by considering the step elastic-field decay along $\uv{z}$, the direction normal to the surface. The step elastic energy ($E^\t{bulk}_\t{step}$) decays with the bulk depth proportionally to $\exp(-h/\xi)$ where $\xi$ is a characteristic length which depends on the step orientation and length. After we verified this relationship we used it to impose the same energy tolerance used for the step-step interaction, \ie, $E_\t{step}^\t{bulk} / U_0 \le 10^{-4}$. The selected bulk depth values are shown in Table~\ref{tab:simulation_box_sizes}. In all simulations presented here the last six layers of $(111)$ planes at the bottom of the simulation box were frozen at their equilibrium position to guarantee that no bending of the structure would occur. The frozen layers were added beyond the values of bulk of depth $h$ shown in Table~\ref{tab:simulation_box_sizes}.

      In order to determine the simulation system dimension corresponding to the step length $L$ it is necessary to consider the assumptions of the CFM, as presented in Sec.~\ref{sec:theory}. This model is not valid for the description of the step at scales smaller than its coarse-graining scale ($\Delta x$), thus we need $L \gg \Delta x$. However, it would be computationally unfeasible to have a step that is excessively large since the relaxation time of normal modes with long wavelength can be very long on MD time scales due the long-range atomic diffusion necessary to change the configuration of these modes. Therefore, it is necessary to study the normal-mode relaxation times before determining what is a satisfactory step length. The details of the analysis of these relaxation times is presented in Sec.~\ref{sec:simulations_error}, and based on the results we have chosen $L \approx 100\Ang$ as shown in Table~\ref{tab:simulation_box_sizes}. This step length is equivalent to the largest relaxation time of the normal mode (with largest wavelength) being $\tau_\t{max} \approx 1 \, \t{ns}$.

    \subsection{\label{sec:simulations_profile} Step profile determination}
      Given any interface between two distinct phases there is no unambiguous approach to determine the interface position from the microscopic atomistic structure of the system \cite{fisher,step_profile_2}. Therefore, there is no algorithm that uniquely defines the step position, \ie, the one-dimensional interface separating two surface terraces. Here we compare two different algorithms \cite{step_profile_3,step_profile_2,step_profile_1,laird_1,laird_2} that determine the step profile from the atomic configurations captured in MD simulations.

\begin{figure}
        \includegraphics[width=0.48\textwidth]{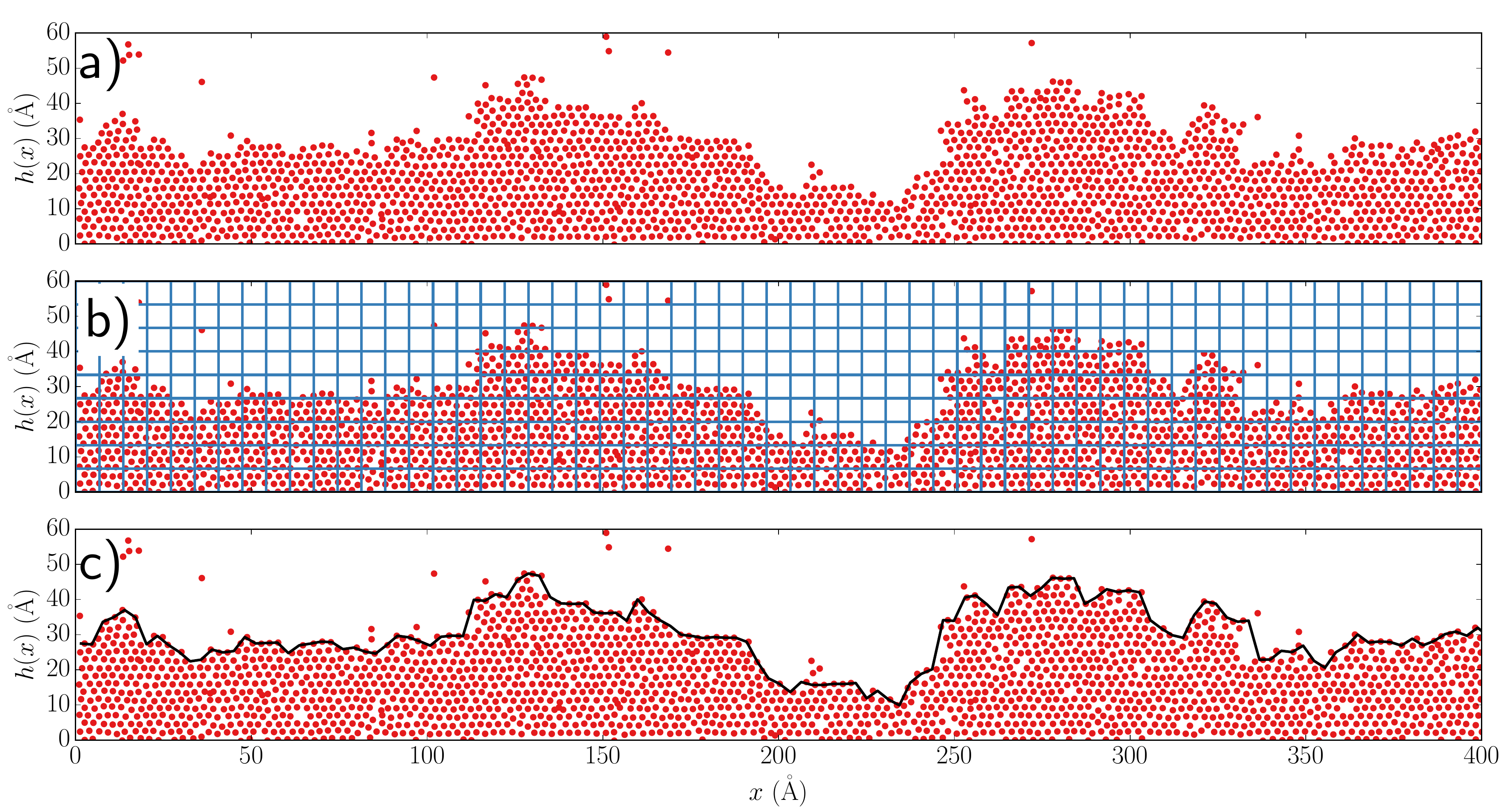}
        \caption{\label{fig:step_profile_bin} Illustration of the grid algorithm for step profile determination. (a) We select the atoms belonging the first surface layer by selecting all atoms with height above some threshold based on the number of atomic layers in the system. (b) Then we define a grid along the $\uv{x}$ direction and divided each of these stripes in equally spaced bins along the $\uv{y}$ direction. (c) The step height is defined as the average height between the first cell with zero density along a stripe and the cell immediately below it.}
      \end{figure}

      Based on the dimensions for $d$ and $h$ presented in Table~\ref{tab:simulation_box_sizes} we have constructed a system with a $[110]$A step with step-step distance $d = 60.5 \Ang$, bulk depth $h = 42.8 \Ang$, and step length $L = 403.6 \Ang$. This system was equilibrated for $10 \ns$ and, afterwards, the step configuration was captured every $0.2 \,\t{ps}$ for $100 \ns$. All results presented in this section were obtained from this simulation. From the MD snapshots the atoms belonging to the top surface layer could be readily identified by counting the number of $(111)$ planes and selecting all atoms with height above some threshold height based on the interplanar separation. One snapshot of the result of this selection is shown in Fig.~\ref{fig:step_profile_bin}(a); from snapshots like this one we want to define the step line profile, being careful to not select any adatom belonging to the surface and also to not accidentally exclude atoms belonging to the step.

      The first algorithm used will be referred to as the ``grid algorithm''. In this algorithm the direction along the step length, $\uv{x}$, is divided in equally spaced bins or delimiting strips, as illustrated in Fig.~\ref{fig:step_profile_bin}(b). We further divide each of these strips along $\uv{y}$, creating rectangular cells, and calculate the density of atoms in each cell. The step height is chosen as the average value of the height of the first bin with zero density and the bin immediately before it. The parameters chosen for the dimension of each bin was $4.5\Ang$ parallel to the step line and $6.7\Ang$ perpendicular to the step line.

\begin{figure}
        \includegraphics[width=0.48\textwidth]{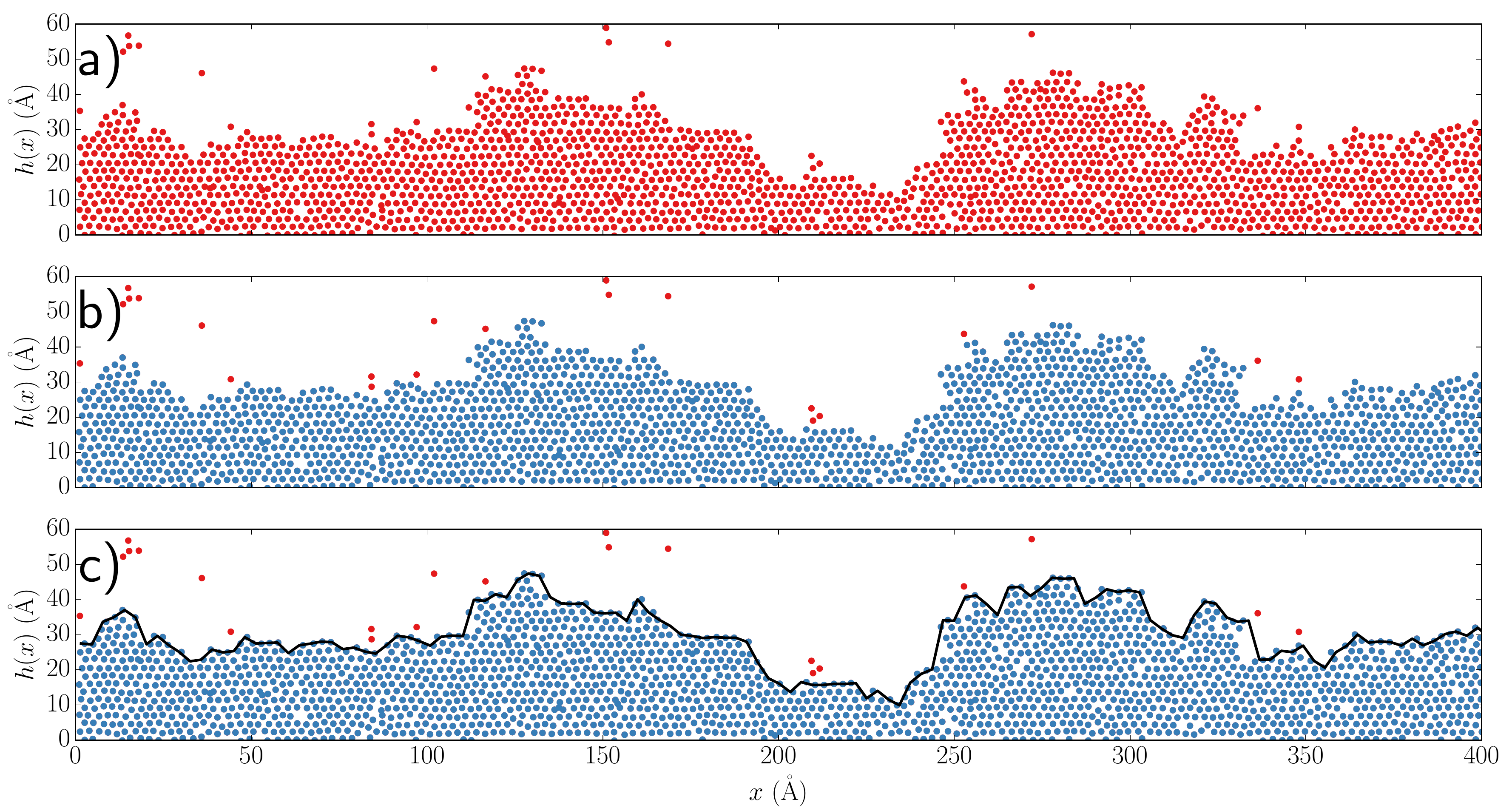}
        \caption{\label{fig:step_profile_cluster} Illustration of the cluster algorithm for step profile determination. (a) We determine the atoms belonging the first surface layer by selecting all atoms with height above some threshold based on the number of atomic layers in the system. (b) Then we find the largest cluster of atoms in that layer (light gray atoms), where we consider two atoms to be neighbors if the distance between them is smaller than some distance $r_\t{max}$. (c) The system is divided in bins along the $\uv{x}$ direction, and the step height in each bin is determined as the atom with the highest value of $y(x)$ inside that bin.}
      \end{figure}

      The second algorithm used is referred to as the ``cluster algorithm'', illustrated in Fig.~\ref{fig:step_profile_cluster}. Once again we start with the atomic configuration of the first layer, then determine all atomic clusters on this layer. Two atoms are considered to belong to the same cluster if there is a path between them through a sequence of neighbor atoms, where we consider two atoms to be neighbors if the distance between them is equal or smaller than some maximum radius $r_\t{max}$. The step atoms are then defined as the largest cluster of atoms, as shown in Fig.~\ref{fig:step_profile_cluster}(b). From the configuration of the atoms belonging to the step the surface can be readily divided into strips along the $\uv{x}$ direction and the atom with the highest value of $y(x)$ within that strip is selected to be the step height at that point. The cluster algorithm has the advantage of having only one adjustable parameter, namely, the maximum nearest-neighbor distance $r_\t{max}$, which can be easily estimated by considerations of the crystal lattice geometry. We have taken $r_\t{max} = 1.2 r_\t{n}$, where $r_n$ is the distance between nearest neighbors in the lattice, and the discretization length along the step line was $2.7 \Ang$.

\begin{figure}
        \includegraphics[width=0.48\textwidth]{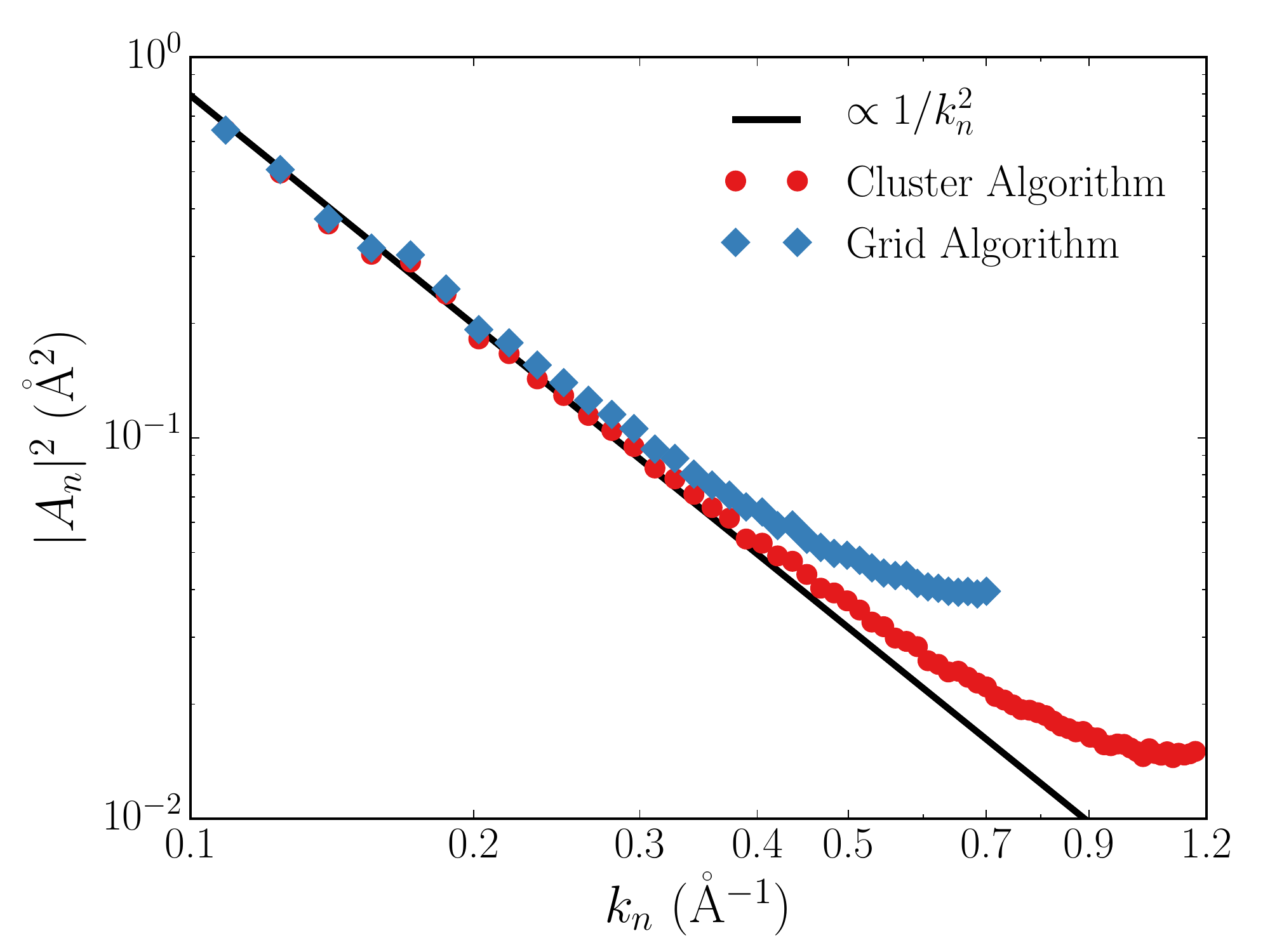}
        \caption{\label{fig:step_profile_algorithm}Comparison of the power spectrum of the Fourier transform of the step line profile obtained using different algorithms to determine the step profile. The cluster algorithm has shown to result in a power spectrum which better follows the $k_n^{-2}$ behavior expected from the capillary-wave model (CFM).}
      \end{figure}

      We have optimized both algorithms with respect to the parameters involved to obtain step profiles that best adjust to the real atomic configurations. Then we performed the Fourier transform of the height profiles and calculated the power spectrum (\ie, $\abs{A_n}^2\, \t{versus} \, k_n$). The comparison of the algorithms is shown in Fig.~\ref{fig:step_profile_algorithm} along with a straight line of slope $-2$. The agreement of the power spectrum with the $k_n^{-2}$ behavior predicted by the CFM in Eq.~\eqref{eq:cwm} is observed, this is an indication that the theory is adequate to describe the step fluctuations at the wavelengths probed in the MD simulations. From Fig.~\ref{fig:step_profile_algorithm} we also see that the long-wavelength modes (small $k_n$) are insensitive to the choice of coarse-graining algorithm, as observed before in this type of analysis \cite{step_profile_1,step_profile_3}. At intermediate values of $k_n$ the cluster algorithm is observed to follow the CFM prediction, Eq.~\eqref{eq:cwm}, to higher wave numbers than the grid algorithm.  Hence, the cluster algorithm is preferred for capturing the step profile details, and this algorithm has been used for analyses of step fluctuations in the remainder of this paper. We also see that for large $k_n$ both algorithms deviate from the $k_n^{-2}$ behavior. This happens because when the normal-mode wavelength becomes comparable to the interatomic distance the atomic vibrations start to interfere with the step oscillations. The CFM was proposed to describe the long-wavelength capillary waves but it does not account for the discrete nature of the atomic configuration and degrees of freedom, thus it is no surprise that when these effects start to become significant (large $k_n$) our results start to deviate from the CFM. We further discuss the validity of the CFM to describe surface steps in Sec.~\ref{sec:results}.

    \subsection{\label{sec:simulations_error} Step normal mode relaxation times}
      Each normal mode in Eq.~\eqref{eq:cwm} has a different relaxation time since mass transport is required for changes in the step configuration. Short wavelength modes can change their configuration quickly since they only require short-range diffusion to modify their amplitudes, while modes with small $k_n$ have long relaxation times that limit the statistics for their sampling in the MD simulations. Thus, these relaxation times ultimately place a limit on the wavelengths that can be probed in the simulations.

\begin{figure}
        \includegraphics[width=0.45\textwidth]{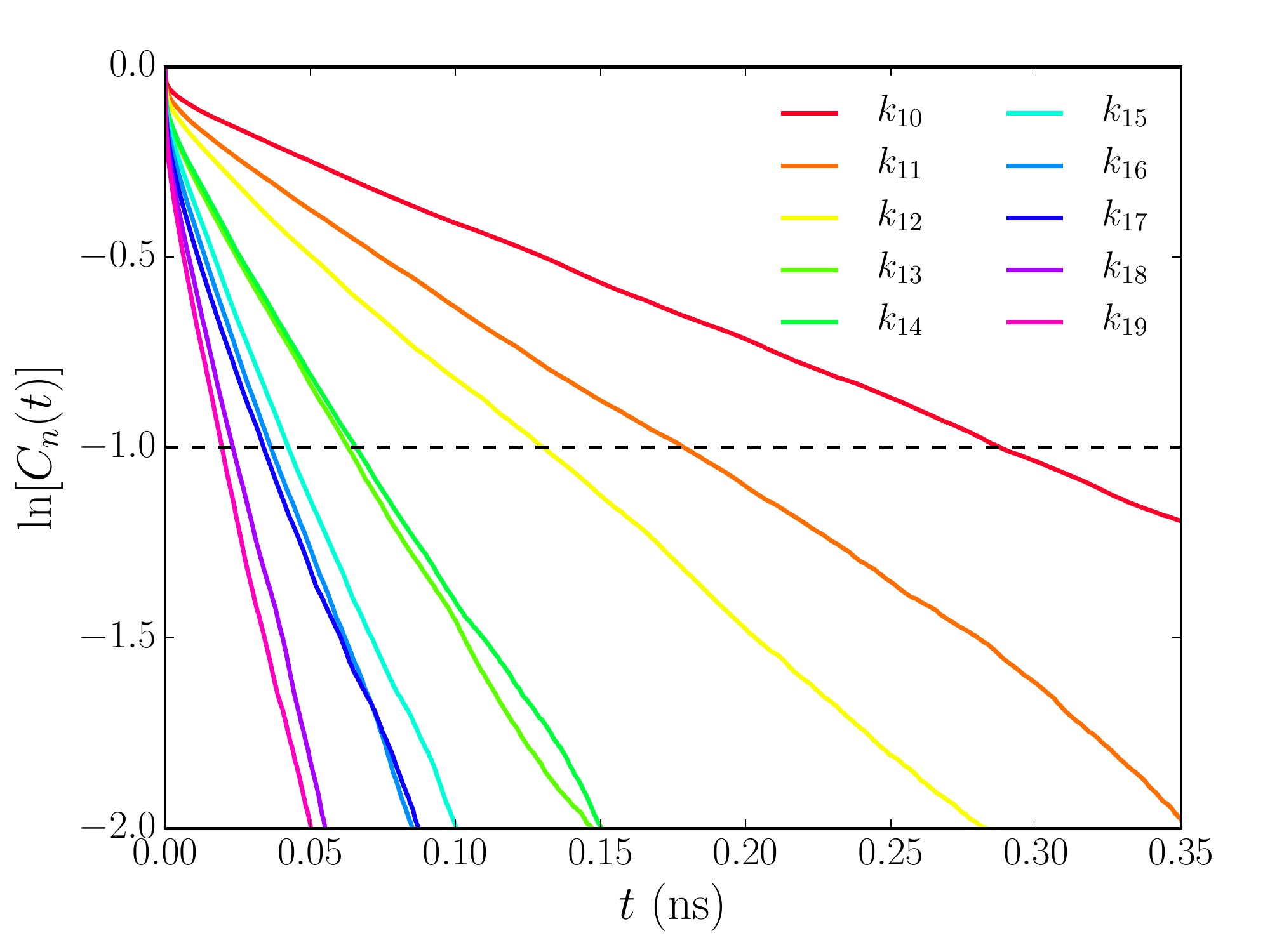}
        \caption{\label{fig:correlation_function_2} Typical autocorrelation functions obtained for the normal modes' amplitudes (normal mode index increases in the clockwise direction or from the upper right to the bottom left). The normal modes relaxation times ($\tau_n$) are obtained by fitting $C_n(t)$ curves to Eq.~\eqref{eq:error}.}
      \end{figure}

\begin{figure}
        \includegraphics[width=0.45\textwidth]{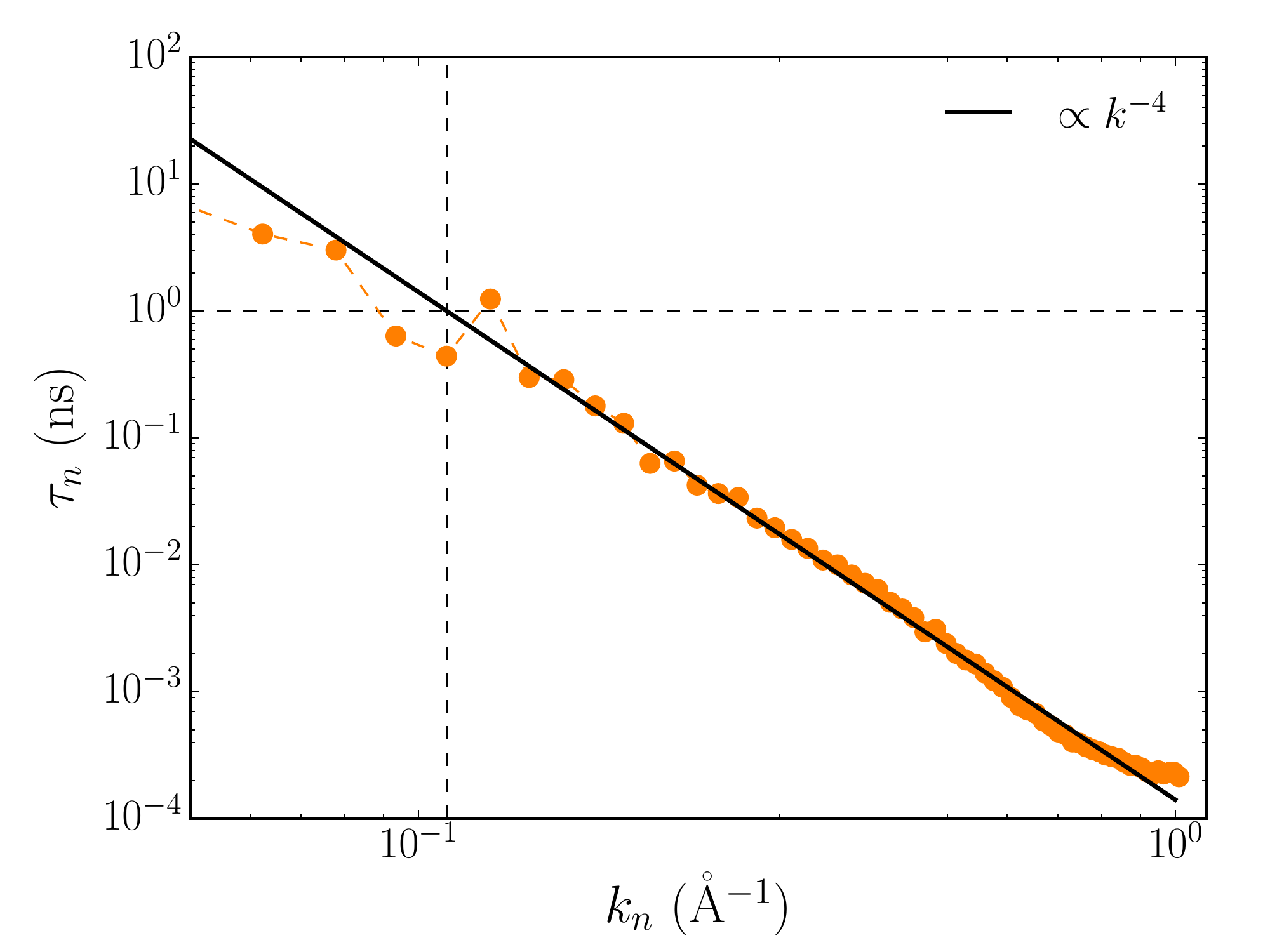}
        \caption{\label{fig:relaxation_time}Relaxation time of each normal mode of the step line profile. $\tau_n$ is obtained by fitting $C_n(t)$ curves to Eq.~\eqref{eq:error}. The horizontal line corresponds to $\tau_n = 1\,\t{ns}$ and the vertical line marks the wavevector value at which the adjusted curve (black solid line) intercepts the $\tau_n = 1 \, \t{ns}$ relaxation time, \ie, the vertical line is at $k_n = k_\t{min}$.}
      \end{figure}

      The relaxation time of the normal modes are analyzed from the simulation data employing a time autocorrelation function of the amplitudes, \ie, if $f_n(t) = \abs{A_n(t)}^2$ then the autocorrelation function $C_n(t)$ is
      \begin{equation}
        C_n(t) = \f{\avg{f_n(t) f_n(0)}}{\avg{f_n(0)^2}} = \exp(-t/\tau_n),
        \label{eq:error}
      \end{equation}
      where $\tau_n$ is the relaxation time of the normal mode of wavevector $k_n$. Representative plots of $C_n(t)$ for selected normal modes are shown in Fig.~\ref{fig:correlation_function_2}, and from such data we can estimate the relaxation time of each mode, as shown in Fig.~\ref{fig:relaxation_time}. The data used to obtain Figs.~\ref{fig:correlation_function_2} and \ref{fig:relaxation_time} was extracted from the simulation performed in Sec.~\ref{sec:simulations_profile} for a $[110]A$ step. 

      Notice in Fig.~\ref{fig:relaxation_time} that the MD simulations resulted in $\tau_n \propto k_n^{-4}$. This result indicates that the step capillary fluctuations are predominantly governed by atomic diffusion \cite{k_scaling_williams} along the step line, as opposed, for example, to diffusion of adatoms on the terrace, which would lead \cite{k_scaling_karma} to $\tau_n \propto k_n^{-3}$, or adatom attachment or detachment to or from the step edge that would be consistent with $\tau_n \propto k_n^{-2}$.

      From the relaxation times obtained in Fig.~\ref{fig:relaxation_time} the step length appropriate for the MD simulation cells is determined as follows. We have limited the maximum relaxation time to be $\tau_\t{max} = 1 \ns$, with this limitation the shortest wavevector we can sample is $k_\t{min} \approx 0.11 \Ang^{-1}$ and the shortest step to contain this wavevector has a length of $L \approx 57 \Ang$. Because we are making conservative choices for the step length of all other orientations, and to increase the number of points used in the fitting of Eq.~\eqref{eq:cwm}, we have chosen to use step lengths of approximately $100 \Ang$. The dimensions for the simulation cells used to obtain the results presented below are listed in Table~\ref{tab:simulation_box_sizes}.

  \section{\label{sec:results} Results}
\begin{figure}
      \includegraphics[width=0.48\textwidth]{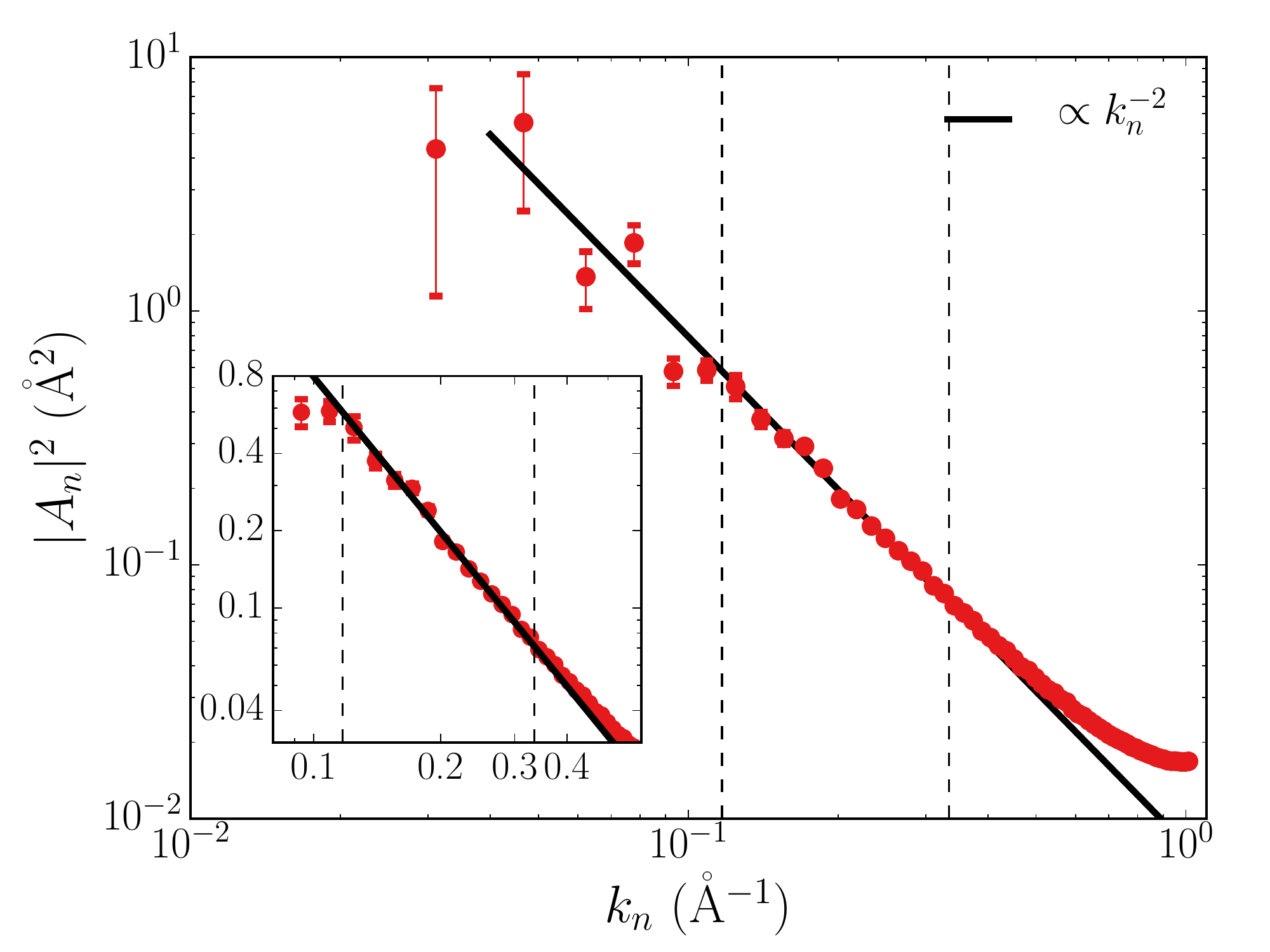}
      \caption{\label{fig:power_spectrum}Power spectrum plot of the Fourier transform of the step line profile. Long wavevectors were discarded since they deviate from the $\avg{\abs{A_n}^2} \propto k_n^{-2}$ behavior due to the presence of atomic vibrations not accounted for the CFM, as presented in Sec.~\ref{sec:theory}. Short wavevectors were discarded because their relaxation time was long enough to be comparable to the total simulation time, hence they cannot be adequately sampled in a MD simulation. The dashed lines delimit the wavelengths used for adjusting the black solid curve to Eq.~\eqref{eq:cwm}. The inset highlights the agreement of the normal modes considered with the CFM predictions.}
    \end{figure}

    Plotted in Fig.~\ref{fig:power_spectrum} are the results of the MD calculated fluctuation amplitudes $\avg{\abs{A_n}^2} \, \t{versus} \, k_n$, obtained as described in Sec.~\ref{sec:simulations_profile} with the cluster algorithm used to characterize the instantaneous step profile. Using the normal modes with wavevectors $k_\t{min} < k_n < k_\t{max}$, where $k_\t{min} = 0.110 \Ang^{-1}$ and $k_\t{max} = 0.335 \Ang^{-1}$ (shown as dotted lines in Fig.~\ref{fig:power_spectrum}), we have verified the $\avg{\abs{A_n}^2} \propto k_n^{-2}$ behavior predicted by Eq.~\eqref{eq:cwm} by adjusting a general power law to these points, as shown in the inset of Fig.~\ref{fig:power_spectrum}.

    As explained in Sec.~\ref{sec:simulations_error} the $k_\t{min}$ value was determined based on the largest relaxation time that can be adequately sampled in the MD simulations we performed. It is clear in Fig.~\ref{fig:power_spectrum} that as $k_n$ decreases below $k_\t{min}$ the error bars become larger due to the reduced sampling statistics, associated with the longer relaxation times and the total simulation time of $100\ns$.

    The value of $k_\t{max}$ was chosen by comparing the MD results in Fig.~\ref{fig:power_spectrum} to a curve with slope $k_n^{-2}$ and visually deciding at which point the data started to diverge from this behavior, $k_\t{max}$ was selected as the average of the $k_n$ value of that point and the one immediately before it. Although the choice of $k_\t{max}$ is not unique (\eg, it depends on the method used to characterize the step profile), it is not completely arbitrary either and reasonable estimates can be made by inspection of the MD results, as shown in Fig.~\ref{fig:power_spectrum}. Specifically, it is clear from Fig.~\ref{fig:power_spectrum} that at large enough $k_n$ the MD data deviates from the $k_n^{-2}$ behavior predicted by the CFM, and the point where this discrepancy becomes statistically significant provides the basis for a reasonable estimate of a lower bound for $k_\t{max}$. Hence, an important criteron is to adjust the $k_n^{-2}$ curve in such a way that the data for small $k_n$ lies accurately on the $k_n^{-2}$ curve. With that reference, the only arbitrariness comes from deciding where the data starts to deviate from the curve location imposed from the small $k_n$ points data. Here the cutoff $k_\t{max} = 0.335 \Ang^{-1}$ corresponds to a wavelength of $\lambda \approx 7.2a$, where $a$ is the atomic distance along the step line for steps along $\Avg{110}$ directions. This result implies that it is necessary to average over approximately seven atoms to eliminate noise due to atomic vibrations and correlated atomic displacements, to correctly capture the expected capillary wave behavior, consistent with the coarse-graining over atomic degrees of freedom necessary to define the step effective Hamiltonian $\mathcal{H}$ described in Sec.~\ref{sec:theory}.

    According to Eq.~\eqref{eq:hf} the step effective Hamiltonian depends on two parameters: the step tension of a straight step ($\sigma$) and the step stiffness ($\tilde{\sigma}$). For each step orientation listed in Table~\ref{tab:simulation_box_sizes} we have run a $100 \ns$ simulation preceded by a $5 \ns$ equilibration period and applied Eq.~\eqref{eq:cwm} to obtain the step stiffness. The result is shown in the first column of Table~\ref{tab:results}. We have employed three different box sizes to test for size convergence: one with twice as much bulk depth and another with twice as much step-step distance (increasing the bulk depth to account for the deeper penetration of the step elastic field). Also listed in Table~\ref{tab:results} are the step energies $U_0$ at $T=0\K$.
    \begin{table}
      \caption{\label{tab:results}Step stiffness for different step orientations and convergence with box size. $d$ is the step-step separation distance, $h$ is the bulk depth, $L$ is the step length, and $U_0$ is the step self energy. The dimensions used are with respect to the box sizes presented in Table \ref{tab:simulation_box_sizes}. The box geometry is illustrated in Fig.~\ref{fig:step_illustration}.}
      \begin{ruledtabular}
        \begin{tabular}{c c c c c}
            \multicolumn{5}{c}{Step stiffness $\tilde{\sigma}$\footnote{All columns have error bar $\pm 1 \, \t{meV/}\Ang$, except for the $U_0$ column where the error bar is $\pm 0.02 \, \t{meV/}\Ang$.} $(\t{meV/}\Ang)$} \\
            & $(d,h,L)$\footnote{The dimensions are given relative to the values shown in Table~\ref{tab:simulation_box_sizes}} & $(d,2h,L)$ & $(2d, 1.7h, L)$ & $U_0$ \\
          \hline
          $[110]$A      & $37$ & $36$ & $37$ & $103.13$ \\
          $[110]$B      & $38$ & $39$ & $37$ & $104.08$ \\
          $[211]\;\;\,$ & $36$ & $36$ & $37$ & $120.53$ \\
        \end{tabular}
      \end{ruledtabular}
    \end{table}

    The zero-temperature values $U_0$ are observed to show a significant anisotropy between the $\avg{110}$ and $\avg{211}$ orientations: a difference of $\approx 14\%$ characterizes these values in Table~\ref{tab:results}. By contrast, at $1300\K$ the MD results yield stiffness values that are isotropic within the statistical precision of the MD data. These results imply that $\tilde{\sigma}\approx\sigma$, \ie, that the contribution of $\sigma''$ to the stiffness is negligible. The observed decrease in anisotropy of the step tension is interpreted to reflect the fact that when the step fluctuation amplitudes become large on the scale of the atomic dimensions the effects of the lattice are averaged out.

  \section{\label{sec:discussion} Discussion}
\begin{figure}
      \includegraphics[width=0.48\textwidth]{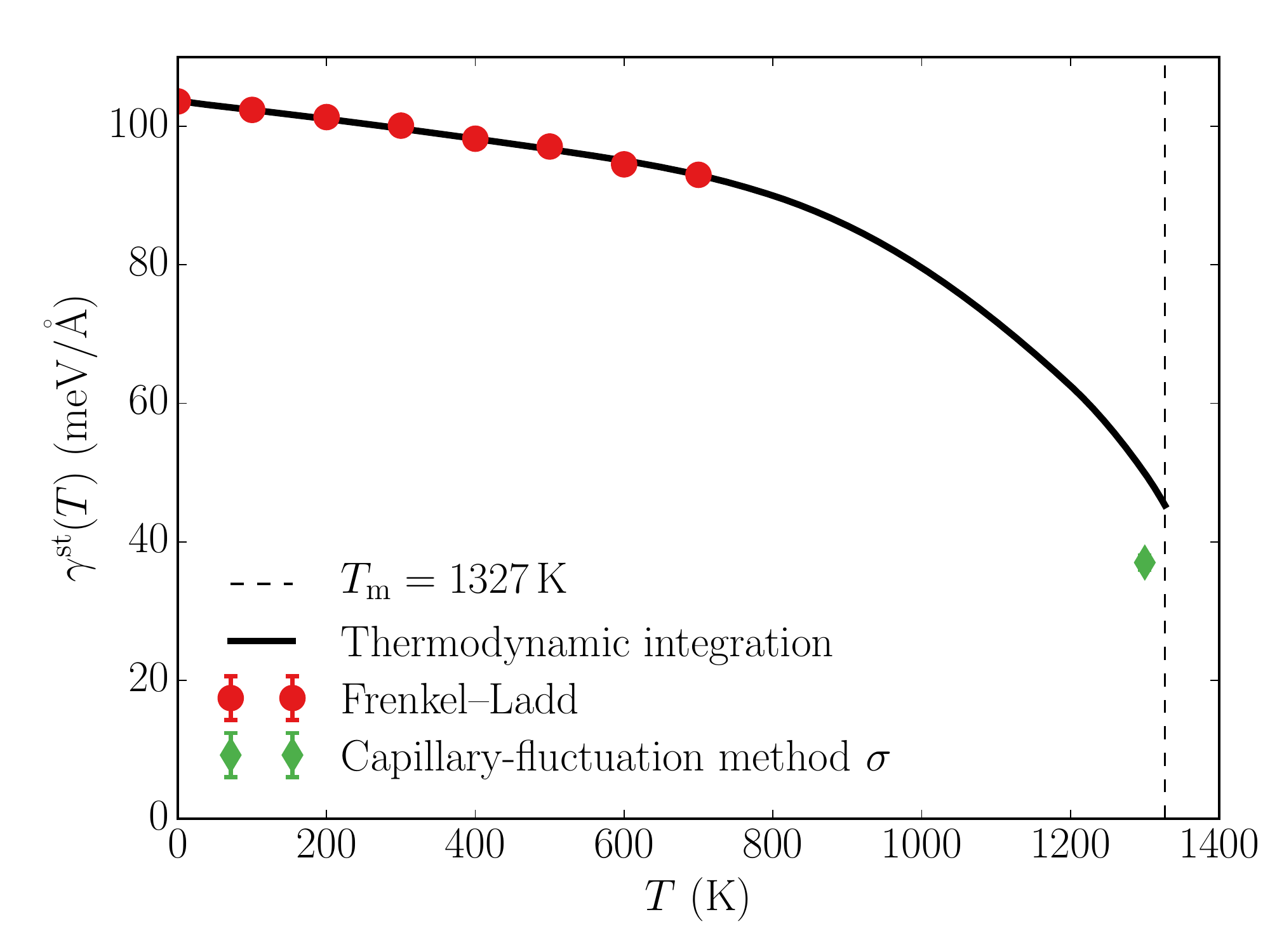}
      \caption{\label{fig:TI}Comparison of the step free energy obtained from thermodynamic integration simulations (Ref.~\cite{freitas_prb}) with the predictions of the capillary-wave model.}
    \end{figure}

    In this section the present CFM results are compared with values of the step free energy for the same system obtained by thermodynamic-integration methods previously \cite{freitas_prb}. In Fig.~\ref{fig:TI} we plot the results from the previous TI calculations, where the solid black line gives the temperature dependence of the step free energy up to the melting point for a $\avg{110}$ step orientation and the red circles are independent results obtained from TI calculations using the Frenkel-Ladd method. Also plotted in Fig.~\ref{fig:TI} with the diamond symbol is the step tension $\sigma$, which is assumed isotropic based on the MD results presented in the previous section, and thus equal to the step stiffness. It can be seen that the value of $\sigma$ obtained from the CFM analysis of the present MD data is lower than the step free energy obtained in Ref.~\cite{freitas_prb} by approximately $25\%$: at $T=1300\K$ the TI results yield $\gamma^\t{st} = (50.4 \pm 0.4) \, \t{meV/}\Ang$, while the CFM yields $\sigma = (37\pm 1) \, \t{meV/}\Ang$. We discuss this discrepancy in what follows in the context of statistical-mechanical theories of capillary fluctuations (\eg, Refs.~\cite{fisher,cwm_original,kayser,abraham1981}).

    We begin by noting the differences in the way the two quantities are defined. In the TI (thermodynamic) formalism, which is Gibbsian in spirit, step free energy is defined as an excess free energy per unit length of the system (simulation block in this case). This thermodynamic formalism does not rely on or characterize the physical length of the fluctuating step. The configurational free energy contributions associated with these fluctuations are naturally included in the TI method. By contrast, in the capillary-wave theory the step tension $\sigma$, introduced in Eq.~\eqref{eq:cwm_hamilt}, is defined as a free energy per unit physical length of the fluctuating step. These considerations alone suggest that the step free energy obtained by the TI method and the step tension defined in the CFM are inherently different. Moreover, in the TI and CFM simulations the average physical length of the step is larger than that of the system dimension along the step by approximately $30\%$ to $35\%$ for the system sizes considered in Ref.~\cite{freitas_prb}. In the TI formalism one could in principle introduce the total excess step free energy per unit of average physical length, which in this case would be approximately $23 \%$ smaller that the value of $\gamma^\t{st}$ given above.

    To formalize the difference between the step stiffness and the step free energy, we note that in the literature, starting with the work of \citet{cwm_original}, there is a distinction drawn between $\sigma$, often referred to as the ``bare'' stiffness, and $\gamma^\t{st}$, the step free energy. In these theories, the latter differs from the former due to the configurational free energy contributions associated with the step fluctuations. Formally, this difference can be derived by the use of Eq.~\eqref{eq:hf} in Eq.~\eqref{eq:F_H}, resulting in an expression for the configurational free energy that depends on the number of modes (coarse-graining length), as well as a length scale that is used in Eq.~\eqref{eq:F_H} to make the partition function dimensionless. These issues are discussed at length by \citet{kayser}, who derives an expression for the configurational contribution to $\gamma^\t{st}$ that depends on system size and geometry. Importantly, this contribution can be shown to be always \textit{negative}, and thus lowers the magnitude of the step free energy ($\gamma^\t{st}$) relative to the bare stiffness ($\sigma$). We note that, as shown in Fig.~\ref{fig:TI}, we find the opposite trend, such that this cannot be the explanation for our finding that $\sigma$ derived from CFM analysis of the MD data is lower than $\gamma^\t{st}$ derived from thermodynamic integration.

    To further consider the origins for this difference, we note that one possibility is that the TI results in Ref.~\cite{freitas_prb} could suffer from hysteresis effects associated with sharp changes in the excess quantities as the step configuration evolves from being straight at low temperatures to rough at higher temperatures. We have investigated these issues in detail in our previous work and concluded that for the system sizes and time scales considered in Ref.~\cite{freitas_prb} the excess quantities behave smoothly and no evidence of artifacts that would bias the TI integration was found.

    We consider then an alternative explanation for the discrepancy between the TI derived value of $\gamma^\t{st}$ and CFM derived value of $\sigma$ shown in Fig.~\ref{fig:TI}. Specifically, as discussed by \citet{fisher} and also noted by \citet{kayser}, the theoretical analysis of \citet{abraham1981} for one-dimensional line interfaces in the 2D Ising model shows that the stiffness that governs the growth the mean-square width of the step with system size is exactly equal to the interfacial free energy. This result suggests a limitation to the classical capillary-fluctuation theory where the two quantities $\sigma$ and $\gamma^\t{st}$ are distinct. This limitation is discussed by Kayser who argues that the stiffness that governs step fluctuations should depend on the wave number $k$ of the fluctuation, \ie, Kayser argues that large-wavelength fluctuations ``see a `renormalized' surface tension'' that differs from the bare stiffness due to the configurational degrees of freedom associated with the smaller wavelength fluctuations. In this picture, it could be possible that the stiffness governing step fluctuations for the lowest wave numbers considered in our simulations, which dominate our fitting of the MD data to extract the stiffness values, is lower than the value of $\gamma^\t{st}$ obtained from TI in Ref.~\cite{freitas_prb} which considered considerably smaller step lengths. These considerations suggest an interesting direction for future work that would involve detailed analysis of the size dependence of both TI and CFM results in the calculation of step free energies.

  \section{\label{sec:summary} Summary and Conclusions}
    Molecular dynamics simulations have been employed in a study of capillary fluctuations of $\Avg{110}$A, $\Avg{110}$B, and $\Avg{211}$ steps on the $(111)$ surface of face-centered cubic copper at a homologous temperature of $0.98$. The simulation results were analyzed within the framework of the statistical-mechanical theory of capillary waves. Specifically, the mean-square fluctuation amplitudes $\avg{\abs{A(k)}^2}$ derived from the simulation data were found to follow the inverse square dependence on wave number ($k^{-2}$) predicted from capillary fluctuation theory over a range $k_\t{min}$ to $k_\t{max}$. This range is bounded by values $k_\t{min}$, below which the fluctuation relaxation times were too long to be adequately sampled in an MD simulation of $100 \,\t{ns}$, and $k_\t{max}$, above which the atomic-scale wavelengths of the fluctuations lead to deviations from from the theoretical $k^{-2}$ scaling. Over this range of wave numbers, the fluctuation relaxation times were observed to display a dependence on wave number consistent with a $k^{-4}$ scaling, corresponding to kinetics limited by step-edge diffusion. We notice that although there are no theoretical restrictions to the application of the CFM to temperatures much lower than $T_\t{m}$, it is possible to find a limit due to computational resources because lower temperatures will require longer simulation times to sample the step fluctuations adequately.

    From the measured fluctuation amplitudes we derive step stiffness values ($\tilde{\sigma}$) for each of the step orientations considered, obtaining values for the largest system sizes of $(37 \pm 1 ) \, \t{meV} / \Ang$ that are isotropic within the statistical precision of the simulation results. The values of $\tilde{\sigma}$ derived by this CFM approach are compared to recent results for the step free energy ($\gamma^\t{st}$) obtained for the same system using an alternative thermodynamic-integration approach \cite{freitas_prb}. The TI values of $\gamma^\t{st}$ and CFM values of $\tilde{\sigma}$ show discrepancies at the level of $25 \%$.
    
    We discuss that the level of discrepancy can be considered within statistical-mechanical theories for step free energies (\eg, Ref.~\cite{fisher}) that draw a distinction between values for the ``bare" stiffness $\sigma$ and the step free energy $\gamma^\t{st}$ that arises from configurational free energy contributions to the latter. The theoretical considerations developed in this previous literature, discussed in the context of the present results, point to an opportunity to use the CFM analysis framework described in this paper and the TI formalism in Ref.~\cite{freitas_prb} to derive more detailed insights into the connection between the (possibly $k$-dependent) values of $\sigma$ that govern fluctuations at intermediate length scales and the step free energy.

  \begin{acknowledgments}
    The authors would like to thank Professors Jeffrey J. Hoyt and Alain Karma for valuable discussions. The authors also gratefully acknowledge the anonymous referee for the comments and details provided about the capillary-wave theory. The research of R.F. and M.A. at UC Berkeley was supported by the U.S. National Science Foundation (Grants Nos. DMR-1105409 and DMR-1507033). R.F. acknowledges additional support from the Livermore Graduate Scholar Program. T.F acknowledges partial support through a postdoctoral fellowship from the Miller Institute for Basic Research in Science at University of California, Berkeley. Additional support for T.F. was provided under the auspices of the U.S. Department of Energy by Lawrence Livermore National Laboratory under Contract DE-AC52-07NA27344.
  \end{acknowledgments}

  \bibliography{bibliography}
\end{document}